\documentclass{article}

\usepackage[utf8]{inputenc}
\usepackage[T1]{fontenc}
\usepackage[margin=1.2in]{geometry}
\usepackage[hidelinks]{hyperref}
\usepackage{url}
\usepackage{tikz}
\usetikzlibrary{calc}
\usepackage{pgfplots}
\pgfplotsset{compat=1.10}
\usepackage{amsmath}
\usepackage{amssymb}
\usepackage{amsfonts}
\usepackage{mathtools}

\definecolor{tab_blue}{HTML}{1F77B4}
\definecolor{tab_orange}{HTML}{FF7F0E}
\definecolor{tab_green}{HTML}{2CA02C}
\definecolor{tab_red}{HTML}{D62728}
\definecolor{tab_purple}{HTML}{9467BD}
\definecolor{tab_brown}{HTML}{8C564B}
\definecolor{tab_pink}{HTML}{E377C2}
\definecolor{tab_gray}{HTML}{7F7F7F}
\definecolor{tab_olive}{HTML}{BCBD22}
\definecolor{tab_cyan}{HTML}{17BECF}

\newcommand{\mess}{\textsf{M.E.S.S.}}

\newcommand{\pymess}{\mbox{\textsf{Py-}\mess{}}}
\newcommand{\pymor}{\textsf{pyMOR}}
\newcommand{\scipy}{\textsf{SciPy}}
\newcommand{\slicot}{\textsf{SLICOT}}
\newcommand{\slycot}{\textsf{Slycot}}
\newcommand{\fenics}{\textsf{FEniCS}}

\newcommand{\bbR}{\ensuremath{\mathbb{R}}}
\newcommand{\Rn}{\ensuremath{\bbR^{n}}}

\newcommand{\Rp}{\ensuremath{\bbR^{p}}}

\newcommand{\Rnn}{\ensuremath{\bbR^{n \times n}}}

\newcommand{\Rpn}{\ensuremath{\bbR^{p \times n}}}

\newcommand{\cH}{\mathcal{H}}
\newcommand{\cL}{\mathcal{L}}

\newcommand{\hA}{\hat{A}}
\newcommand{\hB}{\hat{B}}
\newcommand{\hC}{\hat{C}}
\newcommand{\hE}{\hat{E}}
\newcommand{\hx}{\hat{x}}
\newcommand{\hy}{\hat{y}}

\newcommand{\tran}{^{\operatorname{T}}}
\newcommand{\mtran}{^{-\!\operatorname{T}}}

\newcommand{\Ltwo}{\ensuremath{\cL_{2}}}
\newcommand{\Htwo}{\ensuremath{\cH_{2}}}
\newcommand{\Hinf}{\ensuremath{\cH_{\infty}}}

\newcommand{\p}{\mu}
\newcommand{\pset}{\mathcal{P}}
\newcommand{\pdim}{d}
\newcommand{\trainset}{\mathcal{S}_{\operatorname{train}}}

\DeclarePairedDelimiter{\norm}{\lVert}{\rVert}
\DeclarePairedDelimiterXPP{\normtwo}[1]{}{\lVert}{\rVert}{_{2}}{#1}
\DeclarePairedDelimiterXPP{\normLtwo}[1]{}{\lVert}{\rVert}{_{\Ltwo}}{#1}
\DeclarePairedDelimiterXPP{\normHtwo}[1]{}{\lVert}{\rVert}{_{\Htwo}}{#1}
\DeclarePairedDelimiterXPP{\image}[1]{\operatorname{im}}{\lparen}{\rparen}{}{#1}
\DeclarePairedDelimiterXPP{\myspan}[1]{\operatorname{span}}{\lbrace}{\rbrace}{}{#1}
\DeclarePairedDelimiterXPP{\trace}[1]{\operatorname{tr}}{\lparen}{\rparen}{}{#1}

\let\ge\geqslant%
\let\hat\widehat%

\title{Parametric model order reduction using pyMOR}
\author{%
  Petar Mlinari\'c\thanks{%
    Max Planck Institute for Dynamics of Complex Technical Systems,
    Sandtorstr. 1,
    39106 Magdeburg,
    Germany,
    \texttt{\{mlinaric,saak\}@mpi-magdeburg.mpg.de}%
  } \and
  Stephan Rave\thanks{%
    University of M\"unster,
    Orleans-Ring 10,
    48149 M\"unster,
    Germany,
    \texttt{stephan.rave@uni-muenster.de}%
  } \and
  Jens Saak\footnotemark[1]
}

\begin{document}
\maketitle

\begin{abstract}
  \pymor{} is a free software library for model order reduction
  that includes both reduced basis and system-theoretic methods.
  All methods are implemented
  in terms of abstract vector and operator interfaces,
  which allows direct integration of \pymor{}'s algorithms
  with a wide array of external PDE solvers.
  In this contribution,
  we give a brief overview of the available methods and
  experimentally compare them for the parametric instationary thermal-block
  benchmark defined in~\cite{morRavS20}.
\end{abstract}

\section{Introduction}\label{sec:MRS20:introduction}
\pymor{} is a free software library
for building model order reduction applications
with the Python programming language~\cite{morMilRS16,morpymorweb}.
Originally only implementing reduced basis methods,
since version 0.5,
released in January 2019,
it additionally implements system-theoretic methods such as balanced
truncation~\cite{morMoo81} and IRKA~\cite{morAntBG10}.
Here, we focus on version 2019.2, released in December 2019,
which added support for parametric system-theoretic methods.

We consider model reduction
of the thermal-block model defined in~\cite{morRavS20},
which takes the form
\begin{align*}
  E \dot{x}(t; \p) & = A(\p) x(t; \p) + B u(t), \quad x(0; \p) = 0, \\
  y(t; \p) & = C x(t; \p),
\end{align*}
with system matrices $E, A(\p) \in \Rnn$,
input matrix $B \in \bbR^{n \times 1}$,
output matrix $C \in \Rpn$,
state $x(t) \in \Rn$,
input $u(t) \in \bbR$, and
output $y(t) \in \Rp$,
where $\p \in \pset \subset \bbR^{\pdim}$ is the parameter.
The matrix-valued function $A$ additionally has parameter-affine form
$A(\p) = A_0 + \sum_{i = 1}^{\pdim}{\p_i A_i}$,
where $\p = (\p_1, \p_2, \dots, \p_{\pdim})$.
We also consider a non-parametric version,
for which we write $A$ instead of $A(\p)$.

We begin,
in Section~\ref{sec:MRS20:pymor-philosophy},
with a brief discussion of \pymor{}'s software design.
In Section~\ref{sec:MRS20:mor-methods},
we give a brief overview of the methods implemented in \pymor{}~2019.2.
Next,
we give numerical results in Section~\ref{sec:MRS20:numeric}.
A conclusion follows in Section~\ref{sec:MRS20:conclusion}.

\section{Software design}\label{sec:MRS20:pymor-philosophy}
The central goal of \pymor{}'s design is to allow an easy integration
with external PDE solver libraries.
To this end,
generic interfaces for vectors and operators have been defined
that give \pymor{} access to the solver's internal data structures
representing vectors, matrices or nonlinear operators,
as well as operations on them,
e.g., the computation of inner products or
the solution of linear equation system.

All high-dimensional model reduction operations in \pymor{},
for instance POD computation or Petrov-Galerkin projection,
are expressed in terms of these interfaces.
Compared to a file-based exchange of matrices or solution snapshots,
this approach enables the usage of problem adapted solvers
implemented in the PDE library or
the reduction of very large MPI-distributed problems~\cite{morMilRS16}.

\section{Overview of model order reduction methods}\label{sec:MRS20:mor-methods}
The majority of MOR methods implemented in \pymor{}
are projection-based methods, i.e.,
they consist of finding basis matrices $V$ and $W$ and
defining the reduced-order model as
\begin{align*}
  \hE \dot{\hx}(t; \p)
  & =
    \hA(\mu) \hx(t; \p)
    + \hB u(t), \quad
    \hx(0; \p) = 0, \\
  \hy(t; \p)
  & =
    \hC \hx(t; \p),
\end{align*}
where $\hE = W\tran E V$,
$\hA(\mu) = W\tran A(\p) V = \hA_0 + \sum_{i = 1}^d{\mu_i \hA_i}$,
$\hA_i = W\tran A_i V$,
$\hB = W\tran B$, and
$\hC = C V$.
If $\image{V} = \image{W}$,
we call it a Galerkin projection and otherwise a Petrov-Galerkin projection.

In the following,
we give short descriptions of some projection-based methods
with remarks on their implementation in \pymor{}.

\subsection{Reduced basis method}\label{sec:MRS20:rb}
We consider a weak POD-Greedy algorithm~\cite{morHaa13}
to build a basis matrix $V$
for which the maximum state-space approximation error
\begin{align*}
  \max_{\p \in \trainset}
  \sum_{i = 1}^N \norm{x(t_i; \p) - V \hx(t_i; \p)}_{H_0^1(\Omega)}^2
\end{align*}
for constant input $u \equiv 1$
over some training set $\trainset$ of parameters is minimized
in the Sobolev $H_0^1$-norm.
To this end,
in each iteration of the greedy algorithm
the current reduced-order model is solved
for all $\p \in \trainset$ and
the parameter $\p_{\max}$ is selected
for which an (online-efficient) estimate of the MOR error is
maximized~\cite{morGreP05}.
For this parameter,
the matrix of full-order model (FOM) solution snapshots
\begin{align*}
  X =
  \begin{bmatrix}
    x(t_1; \p_{\max})
    & x(t_2; \p_{\max})
    & \cdots
    & x(t_N; \p_{\max})
  \end{bmatrix},
\end{align*}
is computed,
and the first left-singular vectors of
its $H_0^1$-orthonormal projection onto
the $H_0^1$-orthogonal complement of $\image{V}$
are added to $V$.

Note that,
in the non-parametric case,
POD-Greedy reduces to POD, i.e.,
using the first few left singular vectors of the snapshot matrix $X$ as a
Galerkin projection basis.

\subsection{System-theoretic methods}\label{sec:MRS20:sys-mor}

\subsubsection{Balanced truncation}\label{sec:MRS20:bt}
For non-parametric models,
balanced truncation (BT) consists of solving two Lyapunov equations
\begin{equation}
  \label{eq:MRS20:PQ}
  \begin{aligned}
    A P E\tran
    + E P A\tran
    + B B\tran
    & =
      0, \\
    A\tran Q E
    + E\tran Q A
    + C\tran C
    & =
      0.
  \end{aligned}
\end{equation}
Based on the solutions $P$ and $Q$,
it computes $V$ and $W$ of the Petrov-Galerkin projection.
\pymor{} provides bindings to dense Lyapunov equation solvers in
\scipy{}~\cite{VirGOetal20},
\slycot{}~\cite{Slycotweb} (Python wrappers for \slicot{}~\cite{SLICOTweb}), and
\pymess{}~\cite{messweb}.
For reduction of large-scale models,
there are bindings for low-rank solvers in \pymess{}.
Since \pymess{} does not allow generic vectors,
there is also an implementation of the alternating direction implicit iteration
in \pymor{}~\cite{Bal20}.

It is known that BT preserves asymptotic stability and
has a~priori bounds for Hardy $\Hinf$ and $\Htwo$ errors
depending on the truncated Hankel singular values
(the square roots of the eigenvalues of $E\tran Q E P$).

For parametric models,
there are several possible extensions of
BT~\cite{morBauB09,morWitTKetal16,morSonS17}.
We focus on the simplest global basis approach
by concatenating several local basis matrices.
Let $\p^{(1)}, \p^{(2)}, \dots, \p^{(\ell)} \in \pset$ be parameter samples and
$V^{(1)}, V^{(2)}, \dots, V^{(\ell)}$ and
$W^{(1)}, W^{(2)}, \dots, W^{(\ell)}$
corresponding local basis matrices.
To guarantee asymptotic stability,
we use Galerkin projection with
\begin{align*}
  \begin{bmatrix}
    V^{(1)} & V^{(2)} & \cdots & V^{(\ell)}
    & W^{(1)} & W^{(2)} & \cdots & W^{(\ell)}
  \end{bmatrix}
\end{align*}
after orthogonalization and rank truncation.

\subsubsection{LQG balanced truncation}\label{sec:MRS20:lqgbt}
LQG balanced truncation (LQGBT) is a variant of BT related to the linear
quadratic Gaussian (LQG) optimal control problem.
Unlike BT,
LQGBT consists of solving Riccati equations
\begin{align*}
  A P E\tran
  + E P A\tran
  - E P C\tran C P E\tran
  + B B\tran
  & =
    0, \\
  A\tran Q E
  + E\tran Q A
  - E\tran Q B B\tran Q E
  + C\tran C
  & =
    0.
\end{align*}
Similar to BT, it guarantees preservation of asymptotic stability and has an
a~priori error bound.
As for Lyapunov equations,
\pymor{} provides bindings for external Riccati equation solvers and an
implementation of the low-rank RADI method~\cite{BenBKetal18}.

Additionally,
there is bounded-real BT in \pymor{},
but it currently relies on a dense solver which does not respect the vector and
operator interfaces,
so it is not possible to use it with a PDE solver.

\subsubsection{Iterative rational Krylov algorithm}\label{sec:MRS20:irka}
Iterative rational Krylov algorithm (IRKA) is a locally optimal MOR method in
the Hardy $\Htwo$ norm.
In each step,
it computes (tangential) rational Krylov subspaces
\begin{equation}\label{eq:MRS20:irka-VW}
  \begin{aligned}
    \image{V}
    & =
      \myspan*{
        {(\sigma_1 E - A)}^{-1} B b_1,
        {(\sigma_2 E - A)}^{-1} B b_2,
        \dots,
        {(\sigma_r E - A)}^{-1} B b_r
      }, \\
    \image{W}
    & =
      \myspan*{
        (\sigma_1 E - A)\mtran C\tran c_1,
        (\sigma_2 E - A)\mtran C\tran c_2,
        \dots,
        (\sigma_r E - A)\mtran C\tran c_r
      }.
  \end{aligned}
\end{equation}
The interpolation points $\sigma_1, \sigma_2, \dots, \sigma_r$ for the next step
are chosen as reflected poles $-\lambda_1, -\lambda_2, \dots, -\lambda_r$ of the
projected matrix pencil $\lambda W\tran E V - W\tran A V$
(vectors $b_1, b_2, \dots, b_r$ and $c_1, c_2, \dots, c_r$ are computed based on
the eigenvectors).
Even if the original model has real poles,
the projected poles can be complex.
Since the complex number support is limited in PDE solvers,
solving complex shifted linear systems $(\sigma E - A) x = b$ needs to be done
using an iterative method.
Implementing efficient preconditions for such systems is a future research topic
for \pymor{}.
For this reason,
we demonstrate IRKA only on the non-parametric example in
Section~\ref{sec:MRS20:numeric-0-param}.
In the parametric case,
we only use one-sided IRKA (OS-IRKA),
where $W$ in~\eqref{eq:MRS20:irka-VW} is replaced by $V$,
which guarantees real interpolation points for the heat equation example we
consider.
To generate the global basis matrix,
we concatenate the local basis matrices $V^{(i)}$ and do a rank truncation.

\subsubsection{Generating reduced models}\label{sec:MRS20:sys-mor-reductors}
All system-theoretic methods in \pymor{} can be called similarly.
For instance,
BT can be run with
\begin{verbatim}
bt = BTReductor(fom, mu=mu)
rom = bt.reduce(10)
\end{verbatim}
where \verb|fom| is the (parametric) full-order model (an instance of
\verb|LTIModel|) and
\verb|mu| is the parameter sample.
The \verb|reduce| method of \verb|bt| accepts the reduced order as a parameter
(among others) and
returns the non-parametric reduced-order model \verb|rom| (again an instance of
\verb|LTIModel|).
The basis matrices are then available as \verb|VectorArrays| in \verb|bt.V| and
\verb|bt.W|.

\section{Numerical results}\label{sec:MRS20:numeric}
Here, we present results of applying MOR methods discussed in
Section~\ref{sec:MRS20:mor-methods} to parametric models,
in particular the thermal block example.
To demonstrate \pymor{}'s integration with external PDE solvers,
we used \fenics{}~2019.1.0 (\cite{fenics15}) to define the full-order model.

We use the Hardy $\Htwo$ norm to quantify the results,
which is defined for non-parametric, asymptotically stable systems
\begin{equation}\label{eq:MRS20:fom}
  \begin{aligned}
    E \dot{x}(t) & = A x(t) + B u(t), \quad x(0) = 0, \\
    y(t) & = C x(t),
  \end{aligned}
\end{equation}
as the $\Ltwo$ norm of the impulse response
$h \colon [0, \infty) \to \bbR^{p \times 1}$ defined by
$h(t) = C \exp(t E^{-1} A) E^{-1} B$,
assuming $E$ is invertible~\cite{morAntBG10}.
This can be computed using
\begin{equation}\label{eq:MRS20:h2-norm}
  \norm{h}_{\Ltwo([0, \infty); \bbR^{p \times 1})}^2
  = \trace*{C P C\tran}
  = \trace*{B\tran Q B},
\end{equation}
where $P$ and $Q$ are as in~\eqref{eq:MRS20:PQ}.
Note that for a reduced-order model
\begin{align*}
  \hE \dot{\hx}(t) & = \hA \hx(t) + \hB u(t), \quad \hx(0) = 0, \\
  \hy(t) & = \hC \hx(t),
\end{align*}
the error system
\begin{align*}
  \begin{bmatrix}
    E & 0 \\
    0 & \hE
  \end{bmatrix}
  \begin{bmatrix}
    \dot{x}(t) \\
    \dot{\hx}(t)
  \end{bmatrix}
  & =
    \begin{bmatrix}
      A & 0 \\
      0 & \hA
    \end{bmatrix}
    \begin{bmatrix}
      x(t) \\
      \hx(t)
    \end{bmatrix}
    +
    \begin{bmatrix}
      B \\
      \hB
    \end{bmatrix}
    u(t), \\
  y(t) - \hy(t)
  & =
    \begin{bmatrix}
      C & -\hC
    \end{bmatrix}
    \begin{bmatrix}
      x(t) \\
      \hx(t)
    \end{bmatrix},
\end{align*}
is of the same form as the FOM~\eqref{eq:MRS20:fom},
which allows us to compute $\Htwo$ errors,
i.e., the $\Htwo$ norm of the error system,
using~\eqref{eq:MRS20:h2-norm}.

We chose to use the $\Htwo$ norm because it is independent of the input $u$.
Additionally, it can be computed efficiently using the low-rank Lyapunov
equation solver available in \pymor{}.

We begin with the non-parametric version in
Section~\ref{sec:MRS20:numeric-0-param},
comparing system-theoretic methods with POD\@.
Then, in Sections~\ref{sec:MRS20:numeric-1-param}
and~\ref{sec:MRS20:numeric-4-param} we compare methods for parametric versions.

The source code of the implementations used to compute the presented results can
be obtained from
\begin{center}
  \url{https://doi.org/10.5281/zenodo.3928528}
\end{center}
and is authored by Petar~Mlinari\'c and Stephan~Rave.

\subsection{Non-parametric version}\label{sec:MRS20:numeric-0-param}
\begin{figure}[tb]
  \centering
  \begin{tikzpicture}
    \begin{semilogyaxis}[
        width=0.6\linewidth,
        height=0.4\linewidth,
        xlabel={Reduced order},
        ylabel={Absolute $\Htwo$ error},
        grid=major,
        legend entries={BT, LQGBT, IRKA, OS-IRKA, POD},
        legend style={
          cells={anchor=west},
          legend pos=outer north east,
        },
        mark options={solid},
        cycle list={
          {very thick, tab_blue, mark=*},
          {very thick, tab_orange, mark=x, dashed},
          {very thick, tab_green, mark=triangle*, dashdotted},
          {very thick, tab_red, mark=diamond*},
          {very thick, tab_purple, mark=pentagon*}
        },
      ]
      \foreach \i in {1, 2, 3, 4, 5}
        \addplot table [x index=0, y index=\i] {code/data/zero_param_all.dat};
    \end{semilogyaxis}
  \end{tikzpicture}
  \caption{Comparison of the methods from Section~\ref{sec:MRS20:mor-methods}
    for the non-parametric model (Section~\ref{sec:MRS20:numeric-0-param})}%
  \label{fig:MRS20:zero-param}
\end{figure}
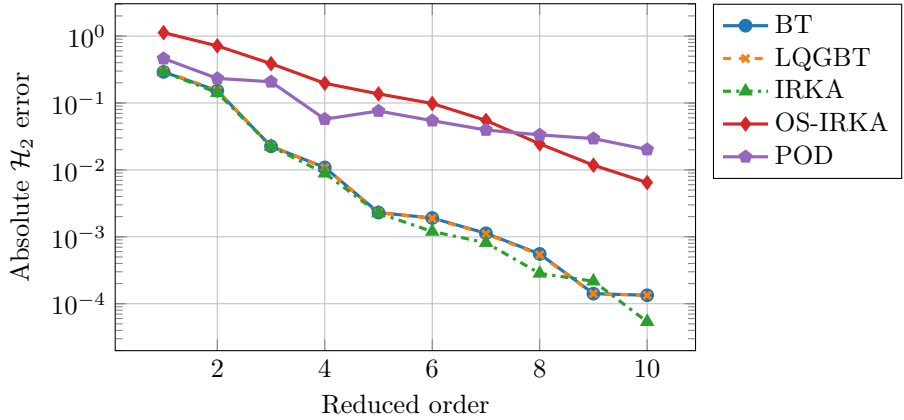

Figure~\ref{fig:MRS20:zero-param} compares BT, LQGBT, IRKA, OS-IRKA, and POD in
terms of $\Htwo$ error.
The POD model was trained using the step response ($u(t) = 1$ for $t \ge 0$).
We see that BT, LQGBT, and IRKA give similar results, while OS-IRKA and
POD give worse errors.
Interestingly, POD is mostly better than OS-IRKA in this example.

\subsection{Single parameter version}\label{sec:MRS20:numeric-1-param}
In this setting, as the training set we chose $10$ logarithmically equi-spaced
parameter values from $10^{-6}$ to $10^2$.
For testing, we added additional $9$ in-between points.
We used BT and OS-IRKA to get reduced models of order $10$ for each parameter
value and concatenated their local bases as explained in
Section~\ref{sec:MRS20:bt}.
After truncation, BT's global basis was of order $175$ and OS-IRKA's was $67$.
To have a fairer comparison,
we further truncated BT's global basis to the same order as OS-IRKA\@.

Figure~\ref{fig:MRS20:one-param-h2-norm} shows the $\Htwo$ norm of the
full-order model for different parameters,
from which we see that it only changes by about an order of magnitude over the
parameter range.
Therefore, we restrict to showing only the absolute $\Htwo$ errors in the
following plots.
In particular,
Figure~\ref{fig:MRS20:one-param-bt-osirka} shows the absolute $\Htwo$ error for
BT and OS-IRKA\@.
Possibly related to BT being a Petrov-Galerkin projection method,
its global basis produces worse results than the local bases.
On the other hand,
OS-IRKA improves with using the global basis.
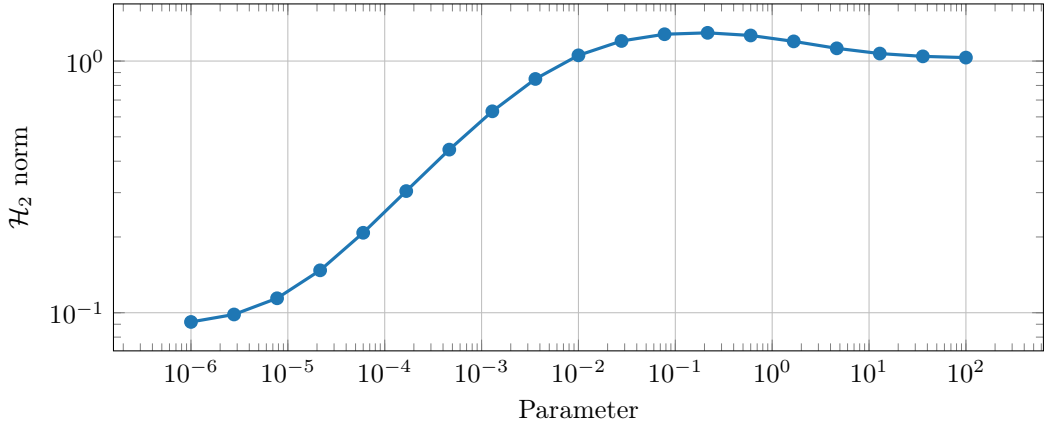
\begin{figure}[tb]
  \centering
  \begin{tikzpicture}
    \begin{loglogaxis}[
        name=abs,
        width=0.9\linewidth,
        height=0.4\linewidth,
        xlabel={Parameter},
        ylabel={$\Htwo$ norm},
        grid=major,
      ]
      \addplot [very thick, tab_blue, mark=*]
        table {code/data/one_param_h2_norm_point.dat};
    \end{loglogaxis}
  \end{tikzpicture}
  \caption{The $\Htwo$ norms of the one-parameter model for different parameter
    values}%
  \label{fig:MRS20:one-param-h2-norm}
\end{figure}
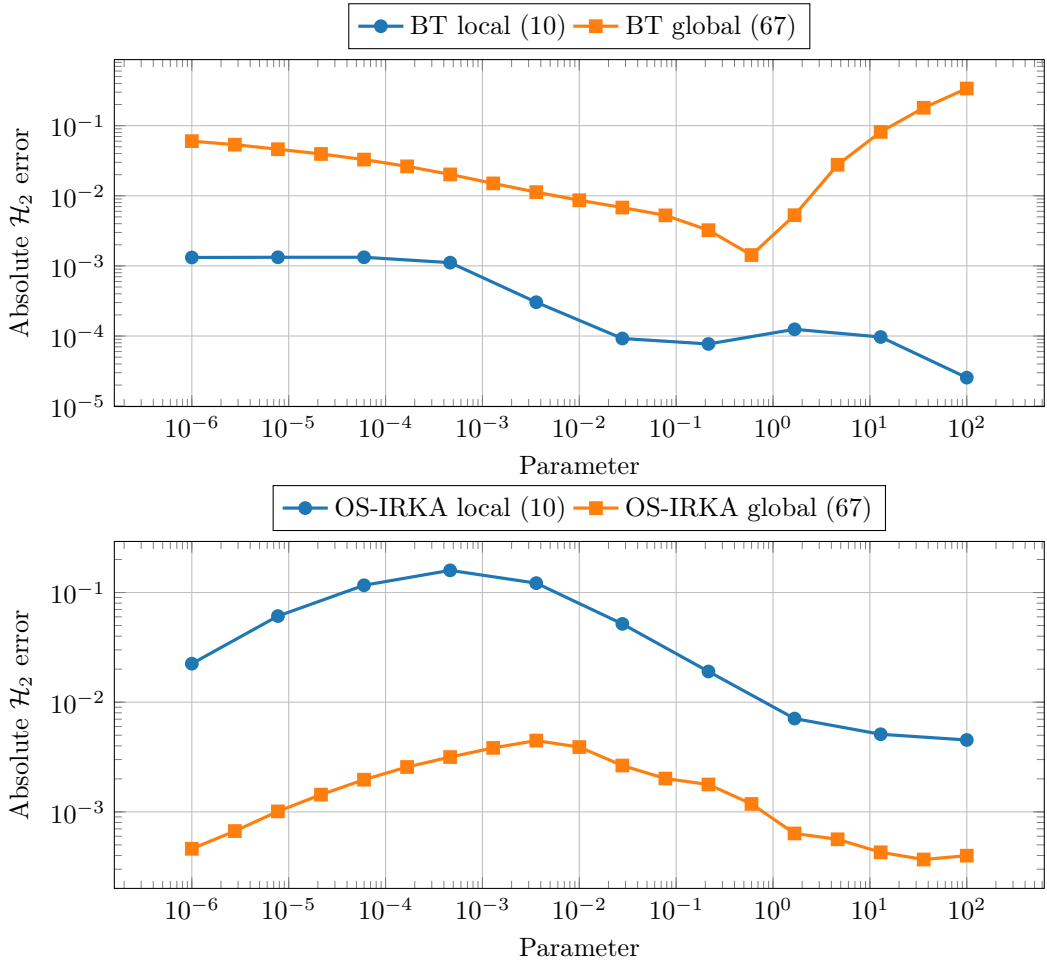
\begin{figure}[tb]
  \centering
  \begin{tikzpicture}
    \begin{loglogaxis}[
        width=0.9\linewidth,
        height=0.4\linewidth,
        xlabel={Parameter},
        ylabel={Absolute $\Htwo$ error},
        grid=major,
        legend entries={BT local (10), BT global (67)},
        legend style={
          at={(0.5, 1.03)},
          anchor=south,
        },
        legend columns=2,
        cycle list={
          {very thick, tab_blue, mark=*},
          {very thick, tab_orange, mark=square*},
        },
      ]
      \addplot table {code/data/one_param_bt_abs_h2_err_local.dat};
      \addplot table {code/data/one_param_bt_abs_h2_err_global.dat};
    \end{loglogaxis}
  \end{tikzpicture}

  \begin{tikzpicture}
    \begin{loglogaxis}[
        width=0.9\linewidth,
        height=0.4\linewidth,
        xlabel={Parameter},
        ylabel={Absolute $\Htwo$ error},
        grid=major,
        legend entries={OS-IRKA local (10), OS-IRKA global (67)},
        legend style={
          at={(0.5, 1.03)},
          anchor=south,
        },
        legend columns=2,
        cycle list={
          {very thick, tab_blue, mark=*},
          {very thick, tab_orange, mark=square*},
        },
      ]
      \addplot table {code/data/one_param_osirka_abs_h2_err_local.dat};
      \addplot table {code/data/one_param_osirka_abs_h2_err_global.dat};
    \end{loglogaxis}
  \end{tikzpicture}
  \caption{Comparison of using local and global bases (see
    Section~\ref{sec:MRS20:bt}) for balanced truncation (BT) and one-sided
    iterative rational Krylov algorithm (OS-IRKA) for the one-parameter model}%
  \label{fig:MRS20:one-param-bt-osirka}
\end{figure}

\begin{figure}[tb]
  \centering
  \begin{tikzpicture}
    \begin{loglogaxis}[
        width=0.9\linewidth,
        height=0.4\linewidth,
        xlabel={Parameter},
        ylabel={Absolute $\Htwo$ error},
        grid=major,
        legend entries={BT (67), OS-IRKA (67), RB (67)},
        legend style={
          at={(0.5, 1.03)},
          anchor=south,
        },
        legend columns=-1,
        cycle list={
          {very thick, tab_blue, mark=*},
          {very thick, tab_orange, mark=square*},
          {very thick, tab_green, mark=triangle*},
        },
      ]
      \addplot table {code/data/one_param_bt_abs_h2_err_global.dat};
      \addplot table {code/data/one_param_osirka_abs_h2_err_global.dat};
      \addplot table {code/data/one_param_rb_abs_h2_err.dat};
    \end{loglogaxis}
  \end{tikzpicture}
  \caption{Comparison of methods for the one-parameter model for fixed reduced
    order (67)}%
  \label{fig:MRS20:one-param-all}
\end{figure}
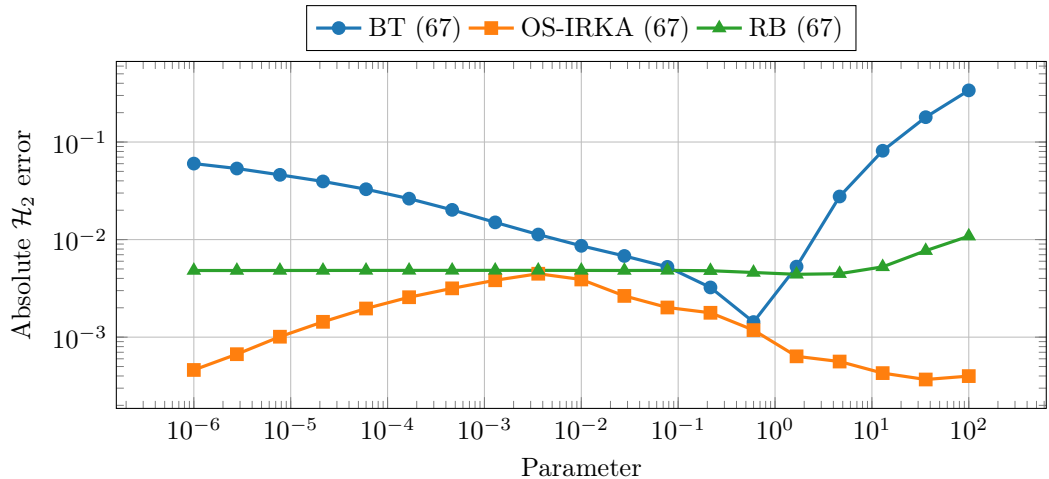

Finally,
Figure~\ref{fig:MRS20:one-param-all} compares BT and OS-IRKA with RB\@.
For RB, we used the same training set to generate a model of order $67$.
In this example,
OS-IRKA performed best near the boundaries of the parameter set and comparable
to other methods in the middle.
On the other hand,
BT gave worst results near the boundaries.
RB produced an almost flat absolute $\Htwo$ error curve,
which is not surprising since it tries to minimize the worst error.

\subsection{Four parameter version}\label{sec:MRS20:numeric-4-param}
\begin{figure}[tb]
  \centering
  \begin{tikzpicture}
    \begin{semilogyaxis}[
        width=0.9\linewidth,
        height=0.4\linewidth,
        xlabel={Parameter index},
        ylabel={Absolute $\Htwo$ error},
        grid=major,
        legend entries={BT local (10), BT global (128)},
        legend style={
          at={(0.5, 1.03)},
          anchor=south,
        },
        legend columns=2,
        cycle list={
          {very thick, tab_blue, mark=*},
          {very thick, tab_orange, mark=square*},
        },
      ]
      \addplot table {code/data/four_param_bt_abs_h2_err_local.dat};
      \addplot table {code/data/four_param_bt_abs_h2_err_global.dat};
    \end{semilogyaxis}
  \end{tikzpicture}

  \begin{tikzpicture}
    \begin{semilogyaxis}[
        width=0.9\linewidth,
        height=0.4\linewidth,
        xlabel={Parameter index},
        ylabel={Absolute $\Htwo$ error},
        grid=major,
        legend entries={OS-IRKA local (10), OS-IRKA global (128)},
        legend style={
          at={(0.5, 1.03)},
          anchor=south,
        },
        legend columns=2,
        cycle list={
          {very thick, tab_blue, mark=*},
          {very thick, tab_orange, mark=square*},
        },
      ]
      \addplot table {code/data/four_param_osirka_abs_h2_err_local.dat};
      \addplot table {code/data/four_param_osirka_abs_h2_err_global.dat};
    \end{semilogyaxis}
  \end{tikzpicture}
  \caption{Comparison of using local and global bases (see
    Section~\ref{sec:MRS20:bt}) for balanced truncation (BT) and one-sided
    iterative rational Krylov algorithm (OS-IRKA) for the four-parameter model.
    The first $20$ parameters are used to construct local bases and global bases
    are tested on further $20$ parameters
    (cf.~Figure~\ref{fig:MRS20:one-param-bt-osirka})}%
  \label{fig:MRS20:four-param-bt-osirka}
\end{figure}
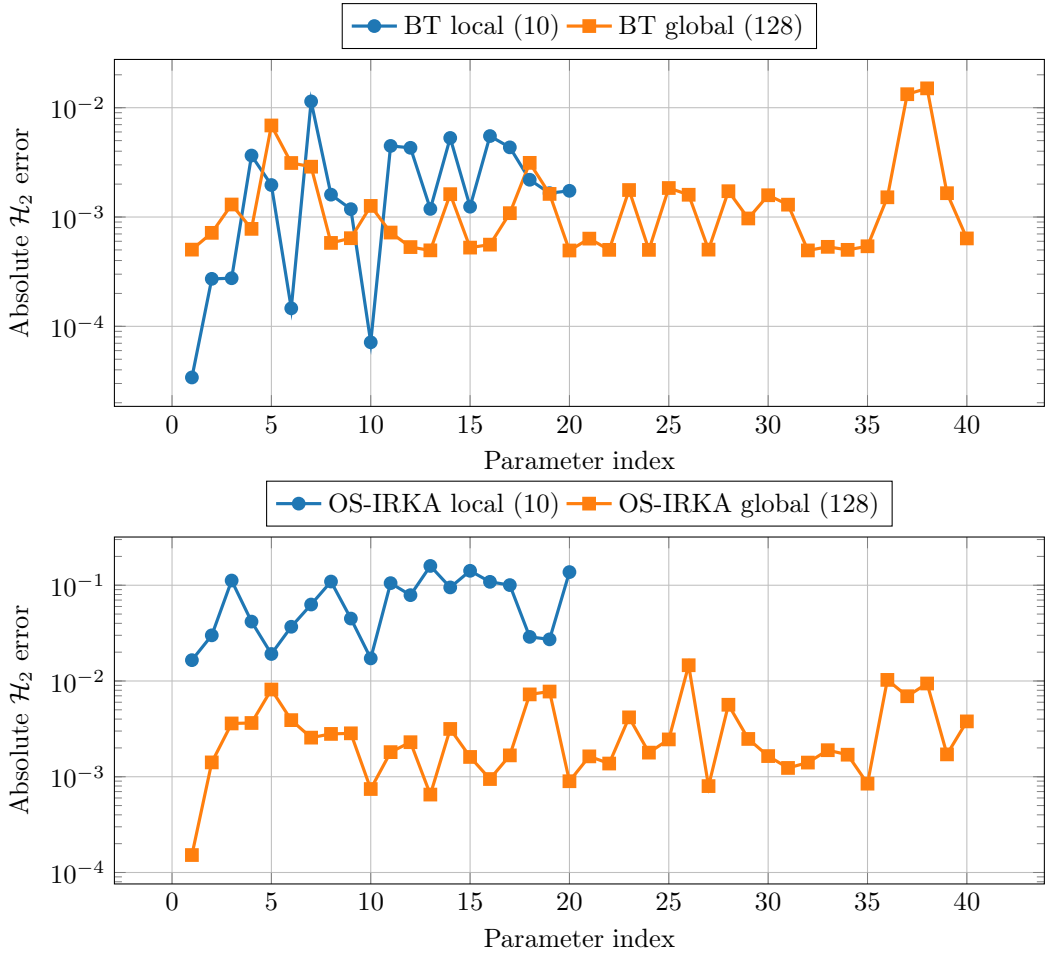

Here, we randomly sampled $20$ points $e_i$ from the uniform distribution over
${[-6, 2]}^4$ to generate the training set $\mu^{(i)} = 10^{e_i}$ and additional
$20$ such points for testing.
As before, we used BT and OS-IRKA to find reduced models of order $10$ at each
training parameter point.
Here, after truncation,
BT's global basis was of order $347$ and OS-IRKA's was $128$.
Figure~\ref{fig:MRS20:four-param-bt-osirka} compares them,
where the first $20$ parameter values are from the training set and the other
for testing.
As we had in the previous example,
OS-IRKA gives better results with the global basis.

\begin{figure}[tb]
  \centering
  \begin{tikzpicture}
    \begin{semilogyaxis}[
        width=0.9\linewidth,
        height=0.4\linewidth,
        xlabel={Parameter index},
        ylabel={Absolute $\Htwo$ error},
        grid=major,
        legend entries={BT (128), OS-IRKA (128), RB (128)},
        legend style={
          at={(0.5, 1.03)},
          anchor=south,
        },
        legend columns=-1,
        cycle list={
          {very thick, tab_blue, mark=*},
          {very thick, tab_orange, mark=square*},
          {very thick, tab_green, mark=triangle*},
        },
      ]
      \addplot table {code/data/four_param_bt_abs_h2_err_global.dat};
      \addplot table {code/data/four_param_osirka_abs_h2_err_global.dat};
      \addplot table {code/data/four_param_rb_abs_h2_err.dat};
    \end{semilogyaxis}
  \end{tikzpicture}
  \caption{Comparison of methods for the four-parameter model for fixed reduced
    order (128)}%
  \label{fig:MRS20:four-param-all}
\end{figure}
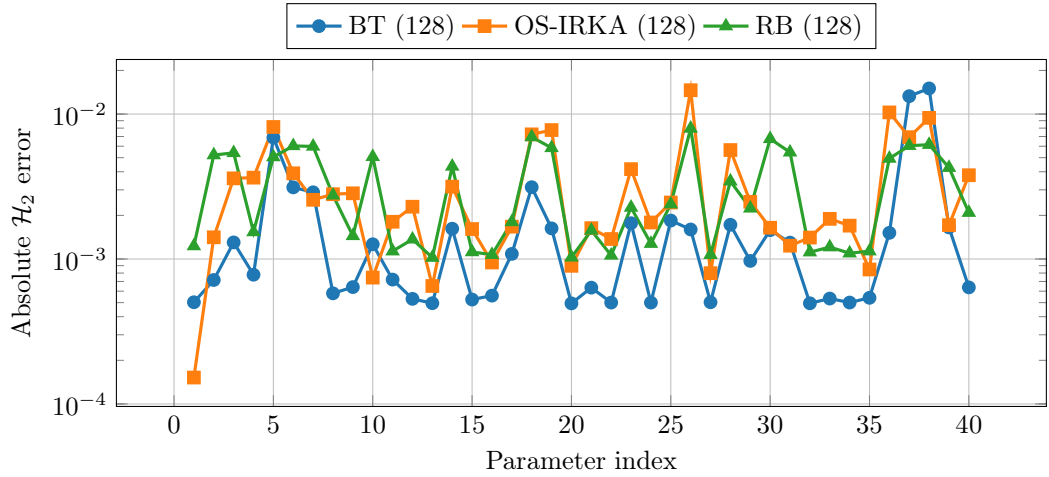

Figure~\ref{fig:MRS20:four-param-all} compares the two methods with RB\@.
We see that they give comparable results,
although they are rather different methods.
On closer inspection,
we note that, in this example,
BT gives better errors the most and
RB shows the smallest maximum error and the least variation in error.

\section{Conclusions}\label{sec:MRS20:conclusion}
We briefly presented \pymor{},
a freely available Python package for MOR,
built on generic interfaces for easy integration with external PDE solvers.
We then described some of the MOR methods implemented in \pymor{},
which includes both system-theoretic and reduced basis methods.
Lastly,
we compared methods on a thermal block benchmark discretized with \fenics{}.

\bibliographystyle{plain}
\bibliography{refs}
\end{document}